\documentclass[twoside]{article}
\usepackage{fleqn,espcrc2}
\usepackage{epsfig}
\title{Residual Chiral Symmetry Breaking in Domain-Wall Fermions}
\author{Chulwoo Jung
\address[UMD]{
Department of Physics,
University of Maryland,
College Park, MD 20742, USA 
{~}}\thanks{
Talk presented by C. Jung.
The authors thank N. Christ and J. Negele for useful
discussions related to the subject of this paper. 
The numerical calculation reported here
were performed on the Calico Alpha Linux Cluster and the QCDSP
at the Jefferson Laboratory, Virginia.
C.J., X.J. and V.G. are supported in part by funds provided by the
U.S.  Department of Energy (D.O.E.) under cooperative agreement
DOE-FG02-93ER-40762. R.G.E. was supported by DOE contract DE-AC05-84ER40150 
under which the Southeastern Universities Research Association (SURA) 
operates the Thomas Jefferson National Accelerator Facility (TJNAF).},
Robert G. Edwards
\address{
Jefferson Lab,
12000 Jefferson Avenue,
MS 12H2,
Newport News, VA 23606, USA 
{~}},
Xiangdong Ji\addressmark[UMD] and
Valeriya Gadiyak\addressmark[UMD]}

\begin{document}

\begin{abstract}
We study the effective quark mass induced by the 
finite separation of the domain walls in the domain-wall 
formulation of chiral fermion as the function of the size of
the fifth dimension ($L_s$), the gauge coupling ($\beta$) and  
the physical volume ($V$). We measure the
mass by calculating the small eigenvalues
of the hermitian domain-wall Dirac operator ($H_{\rm DWF}(m_0=1.8))$ in 
the topologically-nontrivial quenched $SU(3)$ 
gauge configurations. We find that the induced
quark mass is nearly independent of the physical volume, 
decays exponentially as a function of $L_s$, and has 
a strong dependence on the size of quantum fluctuations
controlled by $\beta$. The effect of the choice of 
the lattice gluon action is also studied.
\end{abstract}
\maketitle

Simulating massless or near-massless fermions on a lattice 
is a serious challenge in numerical quantum field theory. 
The origin of the difficulty can be traced to the 
well-known no-go theorem first shown by Nielsen and
Ninomiya, which states that one can not write down a local, 
hermitian, and chirally-symmetric lattice fermion action 
without the fermion doubling problem \cite{nn}. 
Hence to have chirally-symmetric fermions on a lattice, 
one must use nonlocal actions in which the coupling between 
lattice sites do not identically vanish even when the separation
becomes large, as long as one insists on having a 4 dimensional action. 

One of the lattice chiral fermion formalisms that have been studied 
extensively in recent years is the domain-wall fermion, 
first formulated by D. Kaplan \cite{kaplan} and later modified for 
realistic lattice simulation by Shamir \cite{shamir}. In the 
domain-wall construction, one introduces an extra fifth dimension 
$s$ with a finite extension. After discretization, 
the fifth direction
has $L_s$ number of lattice sites. If we put the same
four-dimensional gauge configuration on every four-dimensional 
$s$ slices, the five-dimensional massive theory admits a 
four-dimensional effective theory in which a left-handed chiral 
fermion lives near the $s=0$ slice and a right-handed one near  
$s=L_s-1$. Integrating out the heavy modes 
\cite{neuberger}, one
obtains an effective four-dimensional chiral theory 
in the limit of $L_s\rightarrow \infty$. 
For finite $L_s$, however, the two chiral modes can couple 
to produce an effective quark mass. Strong gauge field 
fluctuations can induce rather strong coupling, and hence rather 
large quark mass. This quark mass is expected to decrease
exponentially as $L_s\rightarrow \infty$ with possible power law 
corrections.

In Ref. \cite{gjj}, we proposed to measure the induced effective
quark mass by considering the eigenvalue of the hermitian 
domain-wall Dirac operator $H_{\rm DWF}(m_0) = R_{s}\gamma_5D_{\rm DWF}(m_0)$, 
where $R_s$ denotes the reflection in the fifth direction.
In the $L_s\rightarrow \infty$ limit, the lattice version of
the Atiyah-Singer theorem \cite{at} guarantees 
the existence of exact
zero modes  in the presence of a instanton background.  
For finite $L_s$, the lowest eigenvalues of the
Dirac operator are not zero. We take the average of these
would-be-zero eigenvalues as the effective quark mass. 
The result is qualitatively consistent with 
those obtained from other methods \cite{cppacs,rbc,aoki}.
Here we report a more systematic study of the effective
quark masses to understand  the effects of different lattice sizes, 
the coupling constant $\beta$, and the form of gluon actions.

\begin{figure}[t]
\begin{center}
\vspace{-0.3in}
\epsfig{file=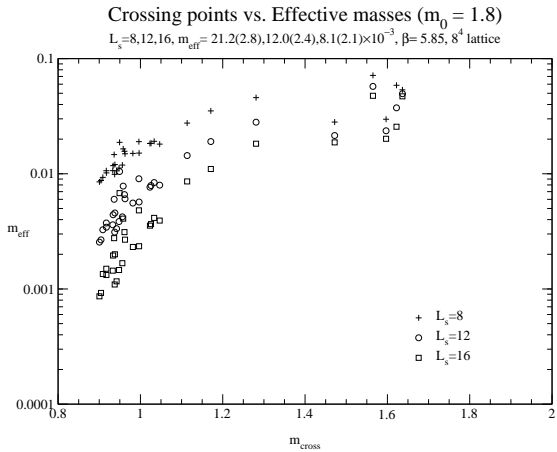,clip=,height=85mm,angle=-90}
\end{center}
\vspace{-0.5in}
\caption{Effective quark mass induced by domain walls with domain wall
height $m_0 = 1.8$ 
for the Monte Carlo configurations at $\beta=5.85, 8^4$ lattice.
$L_s$ is the number of lattice sites in the fifth 
direction.}
\label{fig:585_2_1}
\end{figure}   

To study the effect of the gauge coupling,
we have generated 100 $SU(3)$ 
lattice gauge configurations on a lattice size $8^4$, 
50 each at $\beta =5.85$ and 5.7. We measure the 
spectral flow of the hermitian Wilson-Dirac 
operator to calculate the topological index of the
gauge configurations, and calculate the eigenvalues of 
$H_{\rm DWF}(m_0=1.8)$ corresponding to the nontrivial 
topology of the gauge configurations. 
For larger lattice
spacing, quantum fluctuations are stronger, and some of
these fluctuations can be misidentified as small size 
instantons. It turns out that they can induce strong
couplings between the left-and right-handed chiral modes 
and are detrimental to the existence of the low-energy 
effective theory. Indeed, for the same $L_s$, the effective 
masses are much larger at the smaller $\beta$'s 
than those at $\beta =6.0$. For example, with $L_s=16$, 
$m_{\rm eff}$ is 0.00080(13) at $\beta=6.0$, 0.008(2) at $\beta=5.85$, 
and 0.018(1) at $\beta = 5.7$.  

\begin{figure}[t]
\begin{center}
\vspace{-0.3in}
\epsfig{file=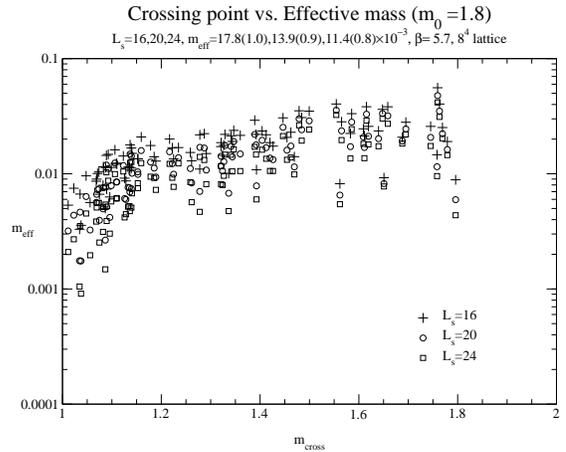,clip=,height=85mm,angle=-90}
\end{center}
\vspace{-0.5in}
\caption{Effective quark mass induced by domain walls with domain wall
height $m_0 = 1.8$ 
for the Monte Carlo configurations at $\beta=5.7, 8^4$ lattice.
$L_s$ is the number of lattice sites in the fifth 
direction.}
\label{fig:57_1}
\end{figure}

Moreover, for the same lattice size, 
physical volume is larger at smaller $\beta$, and hence 
can house more fermionic zero modes. 
As shown in Fig. \ref{fig:585_2_1}
for $\beta=5.85$ and Fig. \ref{fig:57_1} for 
$\beta = 5.70$, the total number of fermionic zero modes 
in the 50 configurations is now 32 and 96, 
respectively. 

For $\beta=5.7$, 
there are several crossings very close to $m_0=1.8$. 
This is in contrast to what has been observed 
at $\beta=6.0$, where the crossings occur mostly
around $m=1.0$ \cite{gjj,specflow2}. 
The equal spacing (in a log plot) between $m_{\rm eff}$ at different 
$L_s$ is a clear signal for the exponential 
decay. However, there is a significant 
variation in the rate among all the crossings, 
as evident in the figures. 
(Note: The recent works of the CP-PACS \cite{cppacs}
 and RBC \cite{rbc} collaborations show
the signal for varying rate of exponential decay and/or
nonvanishing effective mass in the $L_s \rightarrow \infty$ limit.
This behavior is only seen at a much larger $L_s$ than those studied here.
However, it is interesting to note that averaging eigenvalues with 
varying exponential rate can easily reproduce the large $L_s$ behavior of the effective
mass observed in the aforementioned reference.)

To see the volume dependence  
at a fixed $m_0$ and $\beta$, we also measure the effective
mass on a set of 50 configurations on an $8^3\times 16$ lattice 
at $\beta=5.85$. The total number of fermionic zero modes 
is now 64, doubling that on the $8^4$
lattice. 
The average effective
mass turns out to be essentially the same as that 
on the smaller lattice. 

\begin{figure}[t]
\begin{center}
\vspace{-0.3in}
\epsfig{file=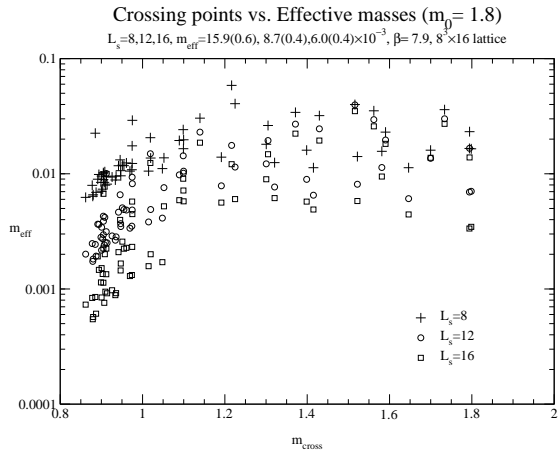,clip=,height=85mm,angle=-90}
\vspace{-0.5in}
\end{center}
\caption{Effective quark mass induced by domain walls with domain wall
height $m_0 = 1.8$ 
for the 50 Monte Carlo configurations generated by the 1-loop, tadpole
improved gauge action \cite{LW} on a $\beta=7.9, 8^3 \times 16$ lattice.
$L_s$ is the number of lattice sites in the fifth 
direction.}
\label{fig:LW}
\end{figure}   

We also measure the effective mass 
on a set of 200 configurations generated from 
the 1-loop L\"uscher-Weisz gauge action \cite{LW} with
the tadpole improvement. (Crossings from only 50 lattices are shown in Fig.
\ref{fig:LW}.) Similar studies using various improved gauge actions,
including  an RG improved action
\cite{iwasaki} are reported in \cite{cppacs,vranas}. 
The gauge coupling ($\beta= 7.9$) corresponds to a 
spacing of $\sim$ 0.16 fm, similar to 
$\beta=5.7$ of the Wilson action. The spectral flow
and the domain-wall eigenvalues are studied with the same 
Wilson-Dirac operator. 
As shown in Fig. \ref{fig:LW}, 
the number of instantons as well as the distribution 
of the crossings differs significantly from the 
Wilson action at similar lattice spacing.
Because of the decrease of small-scale 
quantum fluctuations, the probability of the 
crossing at $m_0 >1.2$ is heavily suppressed, 
and the density of small eigenvalues of $H_{\rm W}(m_0)$ 
is also much smaller ($\sim 2.7 \times 10^{-4}$, compared to 
$\sim 1.8 \times 10^{-3}$ for $\beta=5.7$ Wilson action). 
Therefore, the effective mass 
decreases faster as a function of $L_s$. 
The average effective mass at $L_s=16$ is $\sim 6 \times
10^{-3}$, compared to $18 \times 10^{-3}$ from
the Wilson action at $\beta=5.7$.

\begin{figure}[t]
\begin{center}
\vspace{-0.3in}
\epsfig{file=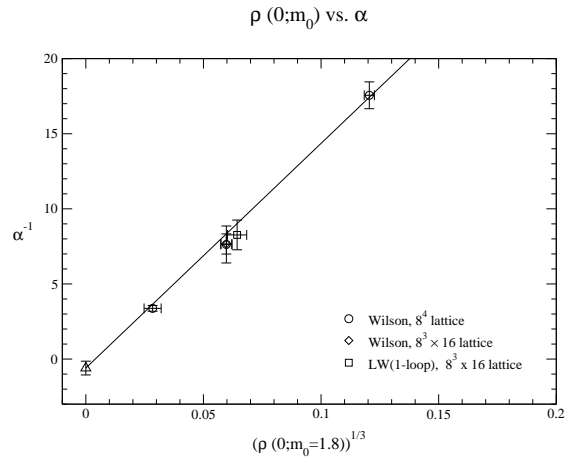,clip=,height=85mm,angle=-90}
\vspace{-0.5in}
\end{center}
\caption{Average coefficient of the exponential decreases
as a function of $\rho(0;m_0)$. The extrapolation of the fit to the
continuum limit for the Wilson gauge backgrounds is shown on the left.
$\rho(0;m_0)$ for the Wilson and improved gauge action are from 
Ref. \cite{specflow2} for $8^3\times 16$ lattices.} 
\label{fig:rho_0}
\end{figure}

This point is also reflected in 
the dependence of the exponential decay rate on the density
of the zero eigenvalues of $H_{\rm W}(m_0)$ ($\rho(0;m_0)$). 
We note that since the gauge fields are replicated along the fifth
dimensional slices, the relevant length scale in the fifth dimension
is the inverse of the rate of exponential decay ($\alpha$) and by simple 
engineering dimensions is given qualitatively by the density of 
zero eigenvalues of $\rho(0;m_0)^{1/3}$. 
In Fig. \ref{fig:rho_0}, we have plotted $1/\alpha$ 
as a function of $\rho(0;m_0)$.
The rate $\alpha$ for each coupling is calculated by fitting 
$m_{\rm eff}(L_s) = m_0\exp[-\alpha L_s]$ at different $L_s$. 
The statistical errors are estimated by doing
correlated fits to single-eliminated jackknife blocks.
The inverse decay rates from all the configurations studied
show an approximate linear scaling as $\rho(0;m_0)^{1/3}$.
This suggests that the density of small eigenvalues is indeed 
the dominating factor for the exponential decay
rate of domain-wall fermions.

\begin{figure}[t]
\begin{center}
\vspace{-0.3in}
\epsfig{file=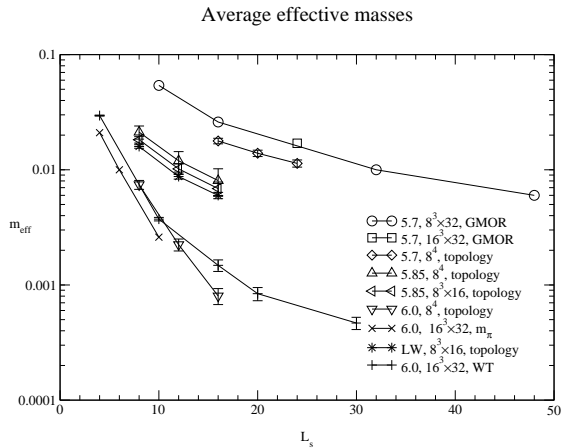,clip=,height=85mm,angle=-90}
\vspace{-0.5in}
\end{center}
\caption{Average effective masses from various observables as a function of
$L_s$. Effective masses from $\beta=5.7$, GMOR relation are from Ref. \cite{fleming}.
Data for $\beta=6.0$, $m_{\pi}$ and the axial Ward--Takahashi(WT) identity are from Ref. \cite{aoki} and
\cite{cppacs}, respectively.
LW denotes the 1-loop, tadpole improved gauge action from
Ref. \cite{LW}.}
\label{fig:meff}
\end{figure} 

Effective masses thus obtained for the different gauge 
coupling and $L_s$ are plotted in Fig. \ref{fig:meff}. 
The data from Ref. \cite{aoki,fleming} are also included 
for comparison. 
For a given gauge coupling, different volumes
and methods of measurements have 
little effect on the size of the effective mass as well as
the rate of exponential decay. However, 
the change of gauge action affects the effective 
mass significantly. 
 It is quite clear from the figure \ref{fig:meff}
that for a practical simulation of the domain-wall
fermion, one either chooses a large $\beta$ with
the conventional Wilson action or an improved
action keeping lattice spacing large.

To summarize, we have studied the residual chiral 
symmetry breaking present in domain-wall fermion by 
measuring the eigenvalues of the hermitian domain-wall 
Dirac operator corresponding to the topology of
the lattice gauge configurations. Individual eigenvalues 
for the topological zero modes show clear exponential
behavior in $L_s$. We regard these eigenvalues as
the induced mass for the surface chiral modes at finite
$L_s$ separation. 

For $L_s$ and $\beta$, we see little variation
of $m_{\rm eff}$ as a function of the volume. This is 
in some sense expected because the coupling of the chiral
modes between the opposite walls has little to do 
with the size of the four-dimensional slice. On the
other hand, a strong dependence on $\beta$ is observed.
In particular, the effective mass is much larger 
at $\beta=5.85$ or 5.7 than that at $\beta=6.0$. 
For the improved gauge action, the 
spurious fluctuations are reduced significantly
 and the $L_s$ needed to obtain a good chiral symmetry
is reduced. Since the additional computation needed for 
the improved action is negligible,
using improved gluon actions may enable us to simulate
domain-wall fermions with larger lattice spacing.

\end{document}